# Flipping of antiferromagnetic to superconducting states in pressurized quasi-one-dimensional manganese-based compounds

Sijin Long[1,2]*, Long Chen[1,2]*, Yuxin Wang[1,2]*, Ying Zhou[1,2]*, Shu Cai[1], Jing Guo[1,4], Yazhou Zhou[1], Ke Yang[3], Sheng Jiang[3], Qi Wu[1], Gang Wang[1,2,4]†, Jiangping Hu[1,2]†, and Liling Sun[1,2,4]†

[1]*Beijing National Laboratory for Condensed Matter Physics, Institute of Physics, Chinese Academy of Sciences, Beijing 100190, China*
[2]*University of Chinese Academy of Sciences, Beijing 100049, China*
[3] *Shanghai Synchrotron Radiation Facilities, Shanghai Institute of Applied Physics, Chinese Academy of Sciences, Shanghai 201204, China*
[4]*Songshan Lake Materials Laboratory, Dongguan, Guangdong 523808, China*

One of the universal features of unconventional superconductors is that the superconducting (SC) state is developed in the proximity of an antiferromagnetic (AFM) state. Understanding the interplay between these two states is one of the key issues to uncover the underlying physics of unconventional SC mechanism. Here, we report a pressure-induced flipping of the AFM state to SC state in the quasi-one-dimensional $A$Mn$_6$Bi$_5$ ($A$ = K, Rb, and Cs) compounds. We find that at a critical pressure the AFM state suddenly disappears at a finite temperature and a SC state simultaneously emerges at a lower temperature without detectable structural changes. Intriguingly, all members of the family present the AFM-SC transition at almost the same critical pressures ($P_c$), though their ambient-pressure unit-cell volumes vary substantially. Our theoretical calculations indicate that the increasing weight of $d_{xz}$ orbital electrons near Fermi energy under the pressure may be the origin of the flipping. These results reveal a diversity of competing nature between the AFM and SC states among the *3d*-transition-metal compounds.

The 3$d$-transition-metal compounds with various magnetic ordered states can often be tuned into an unconventional superconductor by adopting non-thermal control parameters, such as chemical doping, external pressure, and magnetic field. Notable examples include cuprates [1, 2], iron pnictides [3, 4], Cr-based compounds [5], Ni-based compounds [6], and Mn-based binary compounds [7,8], in which the superconductivity develops in the proximity of the magnetic ordered state. How to understand the interplay between the magnetic state and the superconducting (SC) state has been a long-standing challenge for the fields of condensed matter physics and materials sciences [9-16]. Recently, a new class of ternary compounds, $A$Mn$_6$Bi$_5$ ($A$ = Na, K, and Rb) with a quasi-one-dimensional (Q1D) characteristic, has been found. Structurally, these compounds possess a monoclinic structure with the unique [Mn$_6$Bi$_5$]-$A$-[Mn$_6$Bi$_5$] aligning along the $b$ axis [17-19]. In particular, the presence of the antiferromagnetic (AFM) long-range ordered state in their ambient-pressure phase makes this family to be greatly attractive. Application of external pressure on the KMn$_6$Bi$_5$ and RbMn$_6$Bi$_5$ compounds finds that the AFM state is suppressed and then a SC transition is seen [20, 21], which provides a new platform for investigating the correlation between the AFM and SC states. Here, we take the CsMn$_6$Bi$_5$ compound, a new member of this family, as a target material to study the key issue on the correlation, with the aim of obtaining fresh information for a better understanding on the underlying mechanism of unconventional superconductors.

CsMn$_6$Bi$_5$ shares the same Q1D structure motif as $A$Mn$_6$Bi$_5$ ($A$ = Na, K, and Rb), as shown in Fig. 1(a). It crystalizes in a monoclinic space group $C2/m$ (No. 12) with $a$

= 23.6338(14) Å, $b$ = 4.6189(3) Å, $c$ = 13.8948(8) Å, and $\beta$ = 125.4468°(20) [19]. When the electric current is applied either perpendicular or parallel to the $b$ axis, the plot of resistance versus temperature of the sample displays the feature of an AFM transition at $T_N$ = 81 K, and exhibits a larger anisotropic resistivity ratio ($r_{2K} = \rho_\perp/\rho_{//} \sim 50$) at lower temperature (Fig. 1b). The magnetic susceptibility and heat capacity measurements further confirm that the AFM transition occurs at ~ 81 K in our ambient-pressure sample (Fig. 1c and 1d).

First, we performed *in-situ* resistance measurements on the CsMn$_6$Bi$_5$ sample in the pressure range of 2.7-11.6 GPa by using a diamond anvil cell (Fig. 2a). The details about the high-pressure measurements can be found in the Supplementary Information (SI) [22,23]. It is seen that, when the current is applied along the $b$ axis, the resistance as a function of temperature displays a metallic behavior over the experimental temperature range. Upon increasing pressure to 11.6 GPa, a resistance drop is seen at 7.8 K, and a zero-resistance state is observed, as shown more clearly in Fig. 2b. Our results indicate that pressure induces a SC transition of the CsMn$_6$Bi$_5$ compound, in accordance with the results observed in K/RbMn$_6$Bi$_5$ compounds [20, 21]. We repeat the measurements with a new sample (sample #2) cut from different batches and obtain the reproducible results - increasing pressure renders the CsMn$_6$Bi$_5$ compound to be SC at 12.2 GPa (Fig. 2c and 2d). At 14.8 GPa, a zero-resistance state presents (Fig. 2c). The onset SC transition temperature ($T_c$) and the critical pressure ($P_c$) for such a transition is similar to what has been seen in K/RbMn$_6$Bi$_5$ compounds [20, 21]. With further compression to 16.2 GPa, we find the sample investigated loses its zero-resistance state and $T_c$ shifts to lower temperature. At 26 GPa, the resistance drop is not detectable at the temperature down to 1.5 K, the lowest temperature of this study.

To further characterize the pressure-induced superconductivity, we performed the measurements for the smaple#3 and observed the same results - the CsMn$_6$Bi$_5$ compound undergoes a SC transition at 12.6 GPa (Fig. 2f). Application of magnetic field at 13.3 GPa and 17.0 GPa finds that the transition shifts to lower temperature upon increasing the field (Fig. 2g and Fig. S3 in SI [22]). We extracted the field-dependent $T_c$ for the compressed CsMn$_6$Bi$_5$ (Fig. 2h), and estimated the upper critical field at zero temperature by using the Ginzburg-Landau formula [24]: $H_{c2}(T) = H_{c2}(0)[1 - (T/T_c)^2]/[1 + (T/T_c)^2]$. The estimated values of $H_{c2}$ are ~22.9 T at 13.3 GPa and ~23.8 T at 17.0 GPa, which are higher than the value of Pauli limit ($\mu_0 H_p = 1.84 T_c$) [25], implying that the pressure-induced superconductivity in CsMn$_6$Bi$_5$ has unconventional nature [25, 26].

To investigate whether the pressure-induced SC transition is associated with any structural phase transition, we performed high-pressure x-ray diffraction (XRD) measurements for the CsMn$_6$Bi$_5$ sample at the beamline 15W of the Shanghai Synchrotron Radiation Facility. The XRD patterns collected at different pressures are shown in Fig. 3(a). For the pressures ranging from 1.0 GPa to 43.7 GPa, within the pressure range of our transport measurements, we find that all the diffraction peaks can be indexed well by the monoclinic structure in $C2/m$ space group, indicating that no structural transition occurs in the pressure range. The pressure dependences of lattice parameters and cell volume are presented in Fig. 3(b)-(d). It is seen that the lattice parameters $a$, $b$, $c$ and cell volume $V$ decrease monotonously with increasing pressure, while the $β$ angle (the angle between the $a$ axis and the $c$ axis) increases with pressure (Fig. 3e) and saturates above the pressure of 22.5 GPa.

We summarize the high-pressure transport results for the $A$Mn$_6$Bi$_5$ ($A$ = K, Rb, and Cs) family in Fig. 4a, which includes the experimental results of AFM transition temperature ($T_N$) and $T_c$ measured from the compressed K/RbMn$_6$Bi$_5$ [20, 21]. It can be seen that application of pressure continuously suppresses $T_N$ and suddenly induces a SC transition at the border of the AFM order state for all these compounds. Unlike other unconventional superconductors [9, 27, 28], $T_N$ of these compounds terminates at a finite temperature at critical pressure, and $T_c$ appears simultaneously at lower temperature at almost the same pressure point. For our CsMn$_6$Bi$_5$ samples, the measured onset $T_c$ of the sample#1, #2 and #3 appears at 11.6 GPa, 12.2 GPa and 12.6 GPa, respectively, reaches a maximum at 15.6 GPa for the sample#2, 16.2 GPa for the sample#3 and 14.5 GPa for the sample#4 (Fig. S4 in the SI [22]). Upon further compression, $T_c$s decline monotonously and disappear at ~ 27 GPa. The results of $P(T_c)$ obtained from the measurements of CsMn$_6$Bi$_5$ is consistent with what have been seen in K/RbMn$_6$Bi$_5$ [20, 21], suggesting that the all-family members share the same mechanism. It is worth emphasizing that the disappearance of $T_N$ and the appearance of $T_c$ of $A$Mn$_6$Bi$_5$ ($A$ =K, Rb, and Cs) take place at almost the same critical pressure ($P_c$ ~ 12 GPa), and the change of $T_c$ with pressure is also similar. Generally, the different radius of alkali metal ions $A$ ($A$ = K, Rb, and Cs) should deform the lattice differently, which should result in different effects on transport behaviors, such as the different values of $P_c$ for the AFM-SC transition and the different trends of $P(T_c)$. We propose that the observed similar transport results may be attributed to their unique crystal structure, because the $A$Mn$_6$Bi$_5$ ($A$ = K, Rb, and Cs) compounds have almost the same intra-column bond lengths for the [Mn$_6$Bi$_5$]$^-$ columns, a kernel substructure for developing the SC state of this family (see analysis in the SI [22]).

The AFM-SC transition has been found in many compounds with 3$d$ transition metals through chemical doping or applying pressure [4, 5, 7, 9, 29, 30]. The transition can be driven by a structural phase transition [31] and quantum phase transition [4]. However, the AFM-SC transition observed in the $A$Mn$_6$Bi$_5$ compounds is exotic, which has not been seen before. To understand the possible physical origin, we plot the pressure dependence of the $\beta$ angle that represents the lattice deformation degree of the monoclinic phase in Fig. 4b, and find that the value of $\beta$ angle is ~ 125.4° at ambient pressure, but it increases with elevating pressure, reaches to ~126.8° at ~12 GPa where the SC state presents. This result suggests that increasing the $\beta$ angle by ~1.4° may be associated with the suppression of the AFM order state and the emergence of a SC state. At ~ 27 GPa, a transition from a SC state to a non-superconducting state occurs, the $\beta$ angle starts to show a slow decrease with pressure, implying that the $\beta$ angle should be a effective factor in determining the electronic state of these compounds.

To further understand the correlation between the AFM and SC states for this family, we performed first-principles calculations on the electronic structure for the CsMn$_6$Bi$_5$ sample, based on our high-pressure XRD results. The details about the calculations can be found in the SI [22,32-35]. Considering its distinctive Q1D structure, we focused on the percentage of the partial density of state (P-DOS) inside and outside the *ac* plane at the Fermi level. As shown in Fig. 4(c) and Fig. S5 in the SI [22], we find that the P-DOSs contributed by in-plane $d_{xz}$ and out-of-plane $d_{z^2}$ orbital electrons of Mn atoms display an adverse variation upon compression over the experimental pressure range: the P-DOS from $d_{xz}$ orbital electrons shows an increase while the P-DOS from $d_{z^2}$ orbital electrons exhibits a decrease with applying pressure. Significantly, the P-DOSs of the $d_{xz}$ and $d_{z}^{2}$ orbital electrons become nearly identical at the critical pressure (~12 GPa), where the AFM state vanishes and the SC state presents. Upon further compression, we find that the pressure dependent P-DOSs of the $d_{xz}$ and $d_{z}^{2}$ orbital electrons display continuous decrease and increase over the pressure range, respectively. Since the changes of $T_N$ and the P-DOS from the $d_{z}^{2}$ orbital electrons as a function of pressure has the same trend, we propose that the change in the DOS contributed by the $d_{z}^{2}$ orbital electrons is likely associated with the stability of the AFM order state. While an increase of the DOS contributed by the $d_{xz}$ orbital electrons favors the emergence of SC state. Therefore, such pressure-induced changes in the weight of the P-DOS from the $d_{z}^{2}$ and $d_{xz}$ orbital electrons seem to play an important role in the flipping of the AFM and SC states, for which the pressure-induced lattice distortion is considered to be a physical origin.

In summary, we report the first observation of the pressure-induced superconductivity in CsMn$_6$Bi$_5$, and the analysis on the generic flipping of an antiferromagnetic (AFM) state to a superconducting (SC) state in the quasi-one-dimensional $A$Mn$_6$Bi$_5$ ($A$ = K, Rb, and Cs) compounds. We find that these three compounds show an AFM-SC transition at almost the same critical pressure (~ 12 GPa) and present the same trend in $P(T_c)$ with almost the same superconducting dome, indicating that the different alkali metals stacked in the structure of compounds have little effect on the values of $T_c$ and $P_c$. Upon further compression, the superconducting $CsMn_6Bi_5$ converts to a non-superconducting metallic state at ~ 27 GPa and above. High-pressure x-ray diffraction measurements demonstrate that no structural phase transition is found up to 43.7 GPa, except for a continuous decrease in lattice parameters and an increase in $\beta$ angle. Theoretical calculations suggest that the weight of the P-DOSs of the $d_{xz}$ and $d_z^2$ orbital electrons near Fermi energy play an import role for the flipping. These results shed new insight on understandings the correlation between AFM and SC states in unconventional SC materials.

**Acknowledgement**

L. Chen, Y. Zhou, and G. Wang would like to thank Prof. X. L. Chen of Institute of Physics, Chinese Academy of Sciences for helpful discussions. This work was supported by the National Key Research and Development Program of China (Grant Nos. 2021YFA1401800, 2017YFA0302902, and 2018YFE0202602), the National Natural Science Foundation of China (Grant Nos. U2032214, 51832010, 12104487, 12122414, and 12004419), and the Strategic Priority Research Program (B) of the Chinese Academy of Sciences (Grant No. XDB25000000). J. G. and S.C. are grateful for supports from the Youth Innovation Promotion Association of the Chinese Academy of Sciences (2019008) and the China Postdoctoral Science Foundation

(E0BK111).

These authors with star (*) contributed equally to this work.

Correspondence and requests for materials should be addressed to L.S. (llsun@iphy.ac.cn), G.W. (gangwang@iphy.ac.cn), and J.P.H. (jphu@iphy.ac.cn)

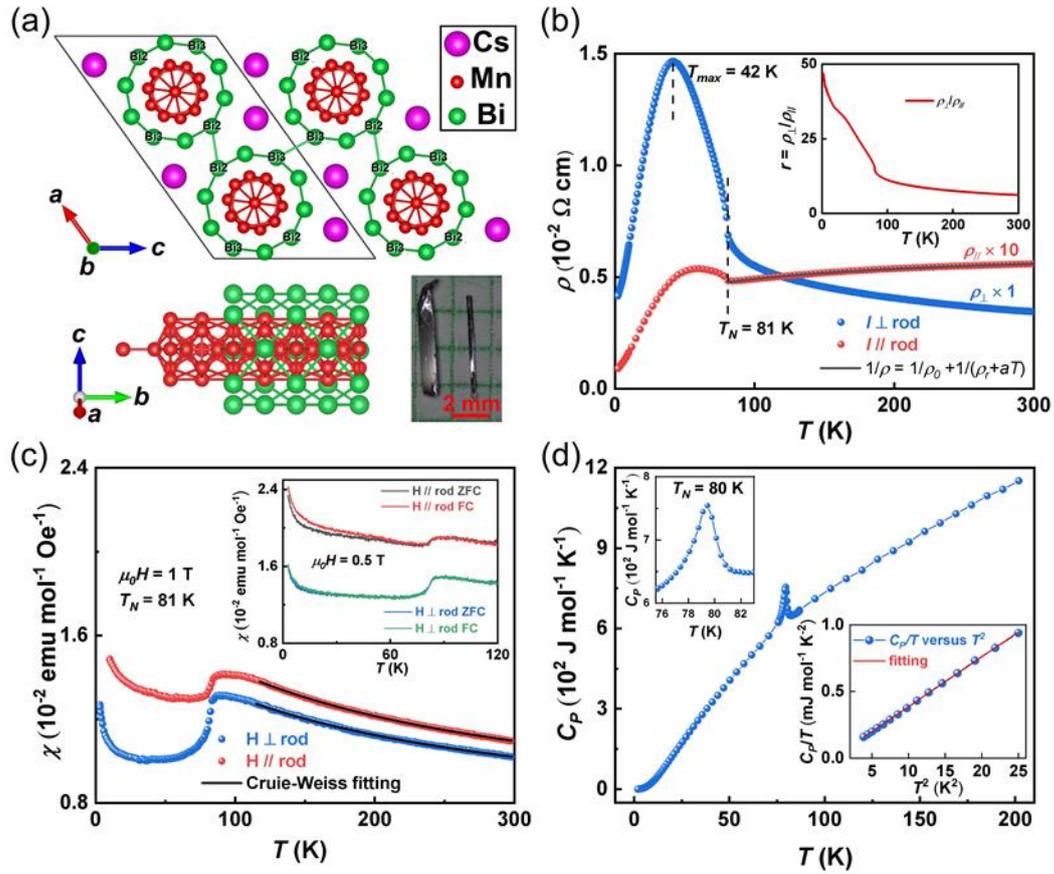

**Figure 1. Crystal structure and properties of CsMn$_6$Bi$_5$ at ambient pressure.** (a) Crystal structure of CsMn$_6$Bi$_5$ viewed along the *b* axis and perpendicular to the *bc* plane. The inter-column bonds Bi2-Bi2 and Bi3-Bi3 are labeled. The lower right shows an optical photograph of the as-grown single crystals with a grid size of 2 mm. (b) Temperature-dependent resistivity $\rho$(T) for CsMn$_6$Bi$_5$ single crystal measured with $I\perp$ rod and $I$//rod ([010] direction). The black solid line is the parallel-resistor formula fit and dashed lines denote the transition temperatures. The inset shows the temperature-dependent anisotropic resistivity ratio $r = \rho_\perp/\rho_{//}$. (c) Temperature-dependent magnetic susceptibility $\chi$(T) of CsMn$_6$Bi$_5$ single crystal under 1 T for $H\perp$ rod and $H$//rod. The black solid lines are corresponding Curie-Weiss fittings from 110 K to 300 K. The inset shows the corresponding ZFC and FC curves under 0.5 T. (d) Temperature-dependent specific heat capacity for CsMn$_6$Bi$_5$ single crystal. The upper inset is the enlarged specific heat capacity around 80 K and the lower inset shows the $C_p/T$ versus $T^2$, where the red solid line is the linear fitting using the Debye model. $T_N$ denotes the AFM ordering temperature determined by the maxima in d$\rho$/dT, d$\chi$T/dT, and $C_p$.

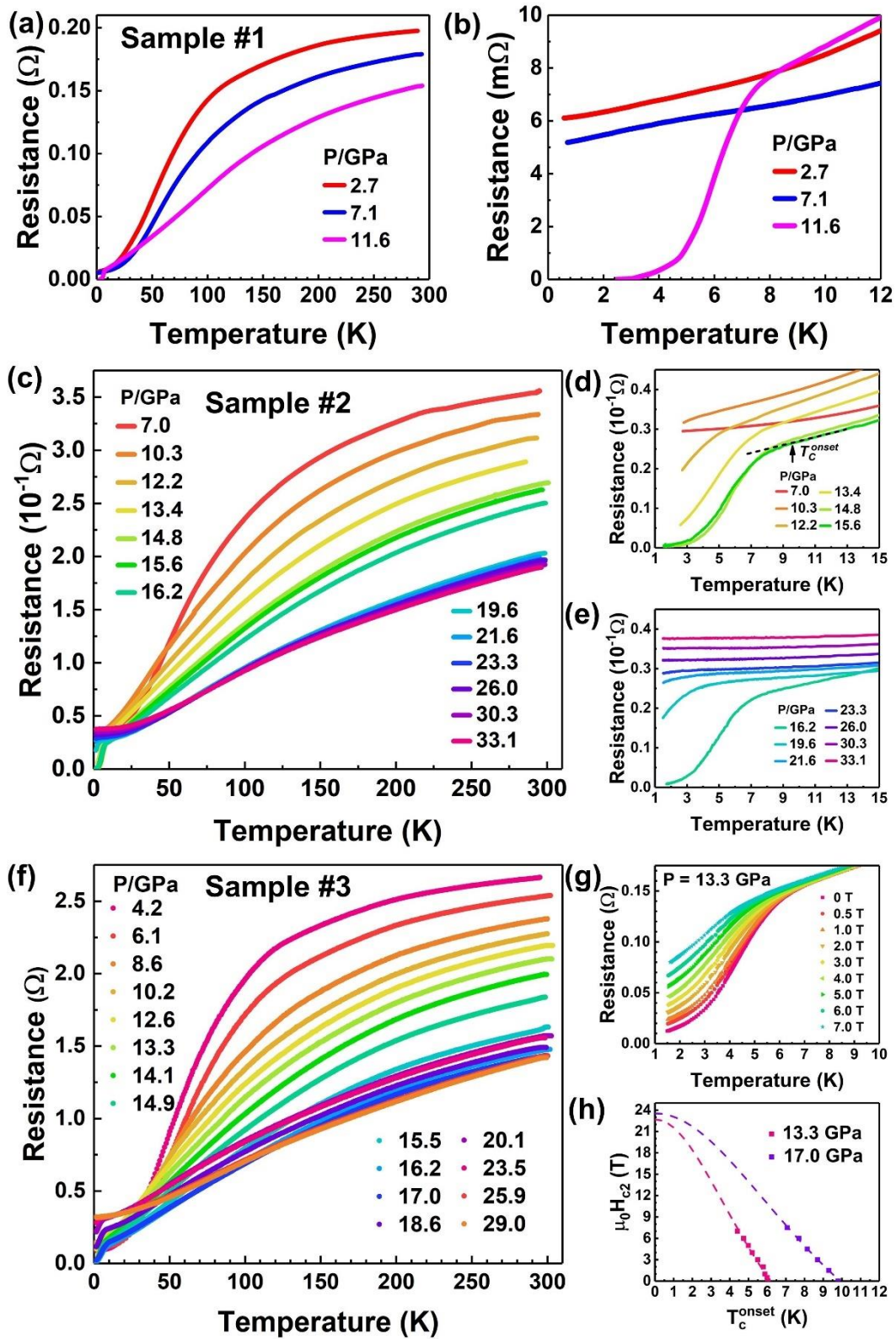

**Figure 2. Temperature dependence of the resistance and determination of upper critical field for the pressurized CsMn$_6$Bi$_5$ samples.** (a) The resistance results

measured from 300 K to 1.5 K for the sample#1 in the pressure range of 2.7-11.6 GPa. (b) The low temperature resistance of the sample#1, showing a clear superconducting transition. (c) The resistance measurements on the sample#2 in the pressure range of 7.0-33.1 GPa. (d) and (e) Enlarged views of the figure c, displaying the details about the superconducting transition of the sample#2. (f) The resistance results of the sample #3 measured from 300 K to 1.5 K in the pressure range of 4.2-29 GPa. (g) The resistance measurements under different magnetic fields for the pressurized sample#3 subjected to 13.3 GPa. (f) Plots of $T_c$ versus upper critical field ($H_{c2}$) for the sample#3 measured at 13.3 GPa and 17.0 GPa, respectively. The dashed lines represent the Ginzburg-Landau (GL) fits to the data of $H_{c2}$.

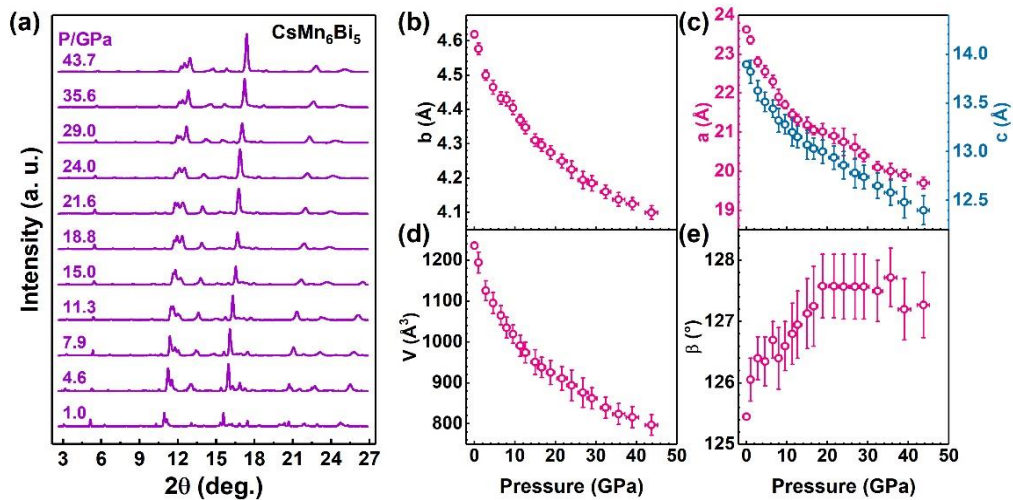

**Figure 3. X-ray diffraction results of the CsMn$_6$Bi$_5$ sample collected at high pressure.** (a) X-ray diffraction patterns measured at different pressures, showing no structural phase transition in the experimental pressure range up to 43.7 GPa. (b) and (c) Lattice parameters *a*, *b*, and *c* versus pressure. (d) and (e) Pressure dependence of cell volume *V* and *β* angle, respectively.

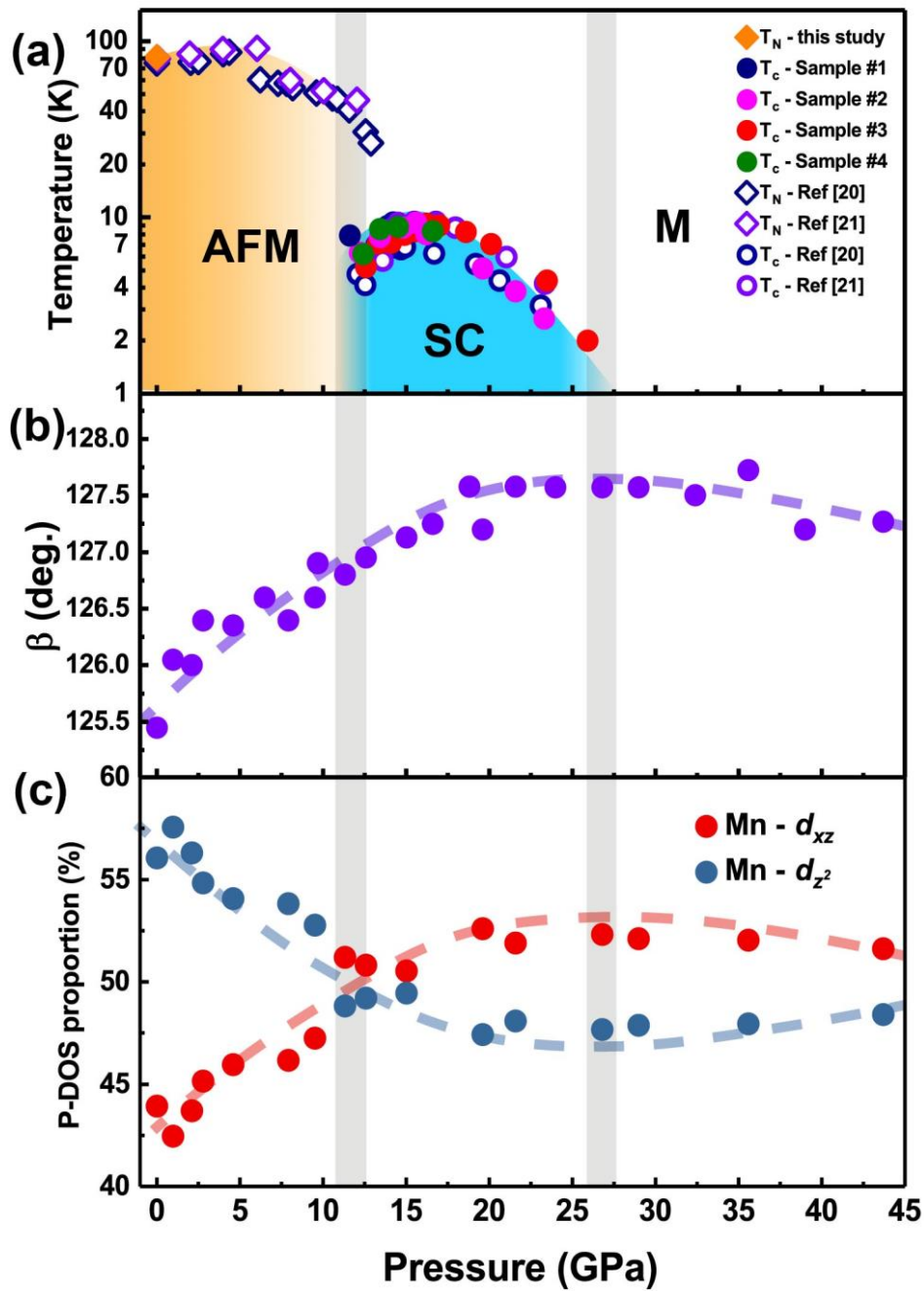

**Figure 4. Pressure-Temperature phase diagram and related information for CsMn$_6$Bi$_5$.** (a) Pressure-Temperature phase diagram, displaying the evolution of antiferromagnetic (AFM) and superconducting (SC) states with increasing pressure. M

represents the non-superconducting metallic state. (b) Pressure dependence of $\beta$ angle. (c) The calculated partial density of state (P-DOS) from the in-plane $d_{xz}$ and out-of-plane $d_{z^2}$ orbital electrons of Mn atoms as a function of pressure. The dash lines are the guides to the eye.